\documentclass[]{JHEP3} 

\preprint{CQUeST-2007-0175}
\usepackage{epsfig,multicol,bbm,colordvi}
\input colordvi

\newcommand\fverb{\setbox\fverbbox=\hbox\bgroup\verb}
\newcommand\fverbdo{\egroup\medskip\noindent%
			\fbox{\unhbox\fverbbox}\ }
\newcommand{\nn}{\nonumber}

\newcommand{\tf}{t_{\rm f}}

\newcommand{\Of}{\Omega_{\rm mf}}
\newcommand{\y}{\mathfrak{a}}

\title{Cosmological Imprint of the Second Law of Thermodynamics
}

\author{Hyeong-Chan Kim 
\\
	Center for Quantum Spacetime, Sogang University,
Seoul 121-742, Republic of Korea,\\
	E-mail: \email{hyeongchan@sogang.ac.kr}}
\author{Jae-Weon Lee \\
	School of Computational Sciences, Korea Institute for Advanced Study, 207-43 Cheongnyangni 2-dong, Dongdaemun-gu, Seoul 130-012, Korea,\\
	E-mail: \email{scikid@kias.re.kr}}
\author{Jungjai Lee\\
	$^3$ Department of Physics, Daejin University,
Pocheon, 487-711, Korea.\\
	E-mail: \email{jjlee@daejin.ac.kr}}

\received{\today} 		
\accepted{\today}		

\date{\today}
\abstract{
We study the evolution of the universe in the presence of inflaton, matter, radiation, and holographic dark energy.
The time evolution of the scale factor is obtained by solving the Friedmann equation of the universe with a good approximation.
We present two independent ways which determine the value of the dark energy constant $d\sim 1$ from the observational data.
The two ways are measuring the deceleration parameter and measuring a universal constant depending only on $d$.  The universal constant is given by a dimensionless combination of three scale factors  at the equipartition times of radiation-matter, radiation-dark energy, and matter-dark energy.
We also discuss that the second law of thermodynamics determines the point of time when the dark energy dominated era begins in the universe.
}

\begin{document}

\section{Introduction}
We show  that the decelerating expansion of the universe with holographic dark energy can not go on forever since it leads to the violation of the second law of thermodynamics eventually.
Interpreting the area of the cosmological event horizon as the entropy of missing information beyond the horizon, we argue that the second law of thermodynamics restricts the time derivative $\dot R_{\rm h}$ to be non-negative ($d\geq 1$), and determines the point of time when the accelerating expansion begins.
To show this, we study the whole history of the universe with the holographic dark energy.

Bekenstein~\cite{bekenstein} formulated the generalized second law which identifies the area of black hole event horizon with its entropy.
The law states that the sum of ordinary entropy and one quarter of the horizon area of the black hole cannot decrease with time.
This identification of  the horizon area as the entropy of the black hole was supported by the presence of thermal Hawking radiation~\cite{hawking} with the black hole temperature $T= \frac{\hbar}{k_B c}\frac{\kappa}{2\pi}$, where $\kappa$ is the surface gravity of the black hole.
In Ref.~\cite{bhinfo,song}, the authors discussed that the horizon area of a black hole denotes the entropy of information erased by the black hole semiclassically by using Landauer's principle~\cite{Landauer}.
It seems natural to relate the area of an event horizon with the quantity of missing information since the horizon, by definition, is the boundary of information.
The information always goes beyond the horizon and never returns.
Therefore, the corresponding entropy must be a non-decreasing function of time, which coincides with the second law of thermodynamics.
In the cosmological case, the de Sitter space has received much attention.
Gibbons and Hawking~\cite{gibbons} have asserted that the generalized second law extends to de Sitter horizons, and detailed investigation~\cite{Davies} confirms this.
The discussion were generalized to quasi-de Sitter spacetime~\cite{Pollock} and to more general cosmological models~\cite{Brustein}.

It is interesting to investigate the role of the thermodynamics in the  Friedmann-Robertson-Walker spacetime.
The spacetime satisfies its first law $dE=T_A dS_A$ with identifying its energy as Misner-Sharp mass~\cite{Misner} at the apparent horizon in various theories of gravity, including the Einstein, Lovelock, nonlinear, and scalar-tensor theories~\cite{Gong}.
This result strongly suggests that the relationship between the first law of thermodynamics of the apparent horizon and the Friedmann equation has a profound physical connection even in the presence of the cosmic microwave background radiation with higher temperature than that of the Hawking temperature of the apparent horizon $T_A$.

The second law of thermodynamics, however, is not guaranteed to be satisfied with the apparent horizon.
Rather, it would be natural to relate the second law with the future event horizon of our universe as in the case of a black hole.
Following the analogy with the black hole case, we may assume that the cosmological horizon area denotes the quantity of missing information.
Since no information returns once it goes out of the event horizon, the quantity of missing information should increase monotonically.
This can be written as $\dot R_{\rm h} \geq 0$, and we interpret this as the second law for the cosmological event horizon.
In relation to the black hole event horizon, we present another argument which support this conclusion:
The area of a black hole horizon never decreases with classical processes.
We may demand that the property also holds for the cosmological event horizons.
In Ref.~\cite{Huang2},  Huang and Li provided several arguments leading to the conclusion $\dot R_{\rm h} \geq 0$. They are the dominant energy condition, the increase of the entropy of the universe, the AdS/CFT correspondence with central charge $d\sim M_p^2/H^2$.
In addition, if $d < 1$, the proper size of the future horizon will
shrink to zero and the IR cut-off will become shorter than the UV cut-off at a finite time
in the future; the very definition of the holographic dark energy breaks down.
In addition, if $d<1$, the size of the Hubble horizon is bigger than the distance to the event horizon $( R_{\rm h}< H^{-1})$ during the inflationary era. In this case, the spectra of the density perturbation based on the Hubble scale may have trouble on the homogeneity.
To avoid these difficulties, in this paper, we assume that the condition $d\geq 1$ holds for our universe.

Therefore, in this case, the entropy bound formulated by Bousso~\cite{Bousso} restricting the total entropy inside to its boundary area is inappropriate since the horizon area is not directly related with the total degrees of freedom of the universe.
The entropy conjecture was tested in adiabatically expanding universe in Ref.~\cite{hongbao}.

We briefly introduce the holographic dark energy model.
Following the idea that the short distance cut-off is related to the infrared cut-off, the holographic dark energy model was first developed by Li~\cite{Li} to explain the present accelerating expansion of the universe. The infrared cut-off relevant to the dark energy was shown to be the size of the cosmological event horizon. The stability of the holographic dark energy under small perturbation was also studied~\cite{myung2,LLW}.
The origin of the holographic dark energy is under investigations.
The entanglement energy on the cosmological event horizon related to the Hawking radiation gives the dark energy of the holographic form~\cite{landDE}.
It was also shown that the spacetime foam uncertainty relation of the form $\delta l \geq l_{\rm p}^\alpha l^{\alpha -1}$ leads to the holographic type energy densities~\cite{Myung}.
The holographic dark energy model was generalized to have an interaction with matter~\cite{Wang} and was constrained by using the supernova data in Refs.~\cite{Huang,Xin}. The second law of thermodynamics for several dark energy models was discussed in Ref.~\cite{setare}.

The Penrose diagram of the universe with a holographic dark energy with equation of state $-1\leq w< -1/3$ was given in Ref.~\cite{chiba}.
\begin{figure}[tbph]
\begin{center}
\begin{tabular}{ll}
\includegraphics[width=.5\linewidth,origin=tl]{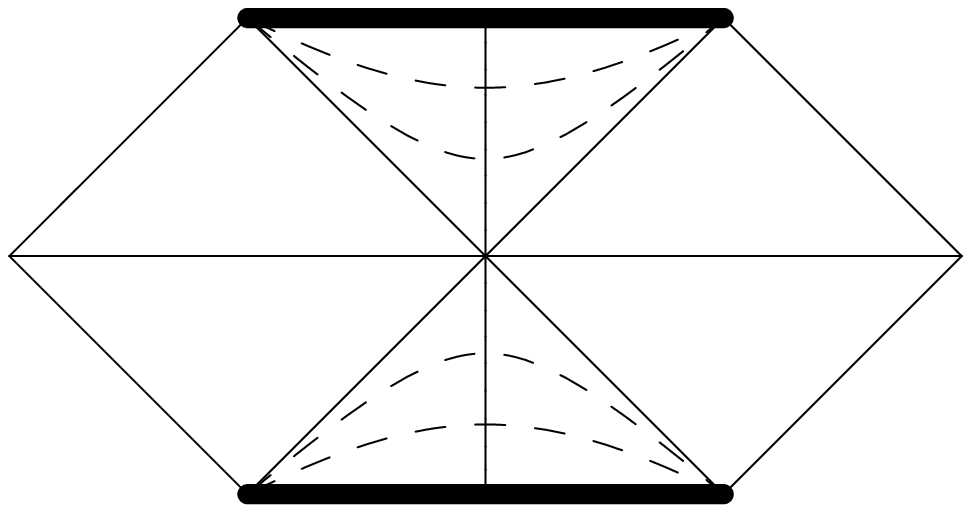} &
\includegraphics[width=.4\linewidth,origin=tl]{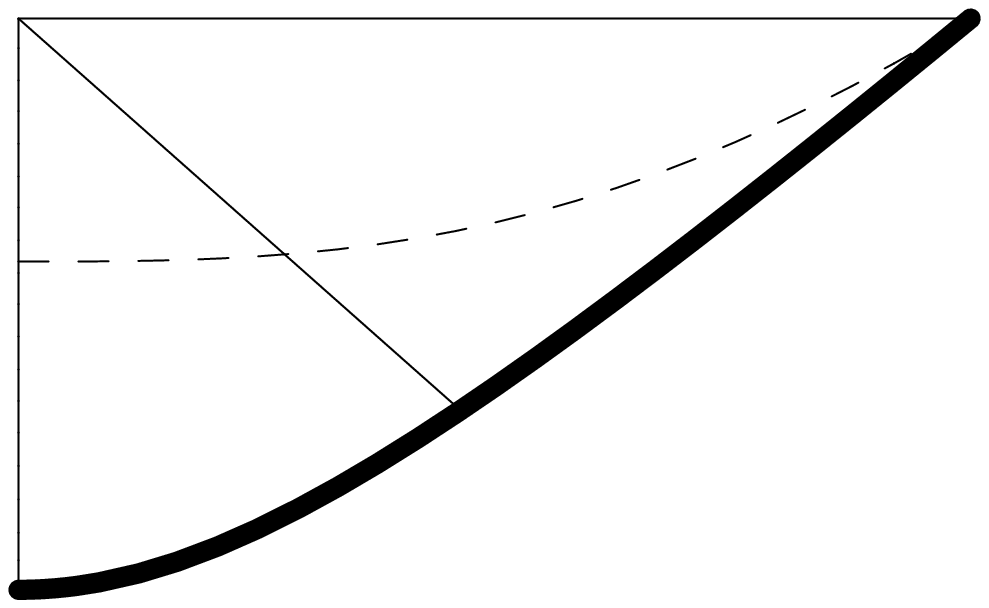}
\end{tabular}
\end{center}
\caption{Penrose diagrams of the Kruskal space and the Robertson-Walker
space with $d\geq 1$.~
The dashed line corresponds to a spacelike slice with a given coordinate time $t$.
 } \label{fig1}
\end{figure}
The event horizon is a surface such that any light departing from the surface cannot arrive at the origin however much time past.
If one compares the two Penrose diagrams in Fig.~1, one may notice that the region we live in is similar to the white hole region of the Kruskal spacetime for the following senses.
First, the singularities are at the past.
Second, the scale factor (or space size for fixed $t$) increases with time.
Third, the asymptotic region ($r\to \infty$) lies behind the horizon.
If one observe the event horizon from the outside of a black hole, one may see a static metric and the black hole entropy counts the quantity of missing information beyond the horizon~\cite{bhinfo}.
In the cosmological case, we live inside the cosmological horizon and its area must be used to count the quantity of missing information beyond the horizon.
Naturally, this entropy (area) is not directly related with the total energy ``inside" the horizon.
Rather, it enumerates how much information of the universe can not be determined from the initial condition of the universe due to the information loss beyond the horizon.
From our viewpoint, the missing information must be related to the dark energy through Landauer's principle~\cite{landDE}, which presents the holographic form of dark energy.
Therefore, considering the second law for the universe, we assume that the cosmological event horizon provides the holographic dark energy.
The presence of the holographic dark energy may change the role of the second law of thermodynamics in the universe.
In this paper, we are interested in the role of the dark energy to the evolution of the universe.

In Sec.~\ref{sec:2}, we construct the precise cosmological model with a holographic dark energy by dividing the evolution of the universe into three phases, the inflation, the consecutive regime of the radiation dominated era (RDE) and the first half of the matter dominated era (MDE), and the consecutive regime of the last half of MDE and the dark energy dominated era (DDE).
In Sec.~\ref{sec:3}, we describe the evolution of the physical quantities in detail for each phase.
In Sec.~\ref{sec:4}, we summarize the results and discuss the physical role of the second law of thermodynamics.

\section{Cosmological model with the holographic dark energy}
\label{sec:2}
In this paper, we consider the flat ($k=0$) Friedmann universe which is favored by observations~\cite{spergel} and inflationary theory~\cite{bassett}.
The model is described by the metric
\begin{eqnarray*}
ds^2=-dt^2+a^2(t)d\Omega_{(3)}^2,
\end{eqnarray*}
where $a(t)$ is the scale factor as usual.
We assume that there are four different kinds of energy densities in our universe denoted by the
inflaton $\rho_{\rm inflaton}$, the radiation $\rho_{\rm r}$, the matter $\rho_{\rm m}$, and the holographic dark energy $\rho_{\rm h}$.
Each energy density except for the inflaton has specific form of behaviors on the scale factor $a(t)$ as
\begin{eqnarray} \label{rho:t}
\rho_{\rm r}(t)= \rho_{r0}\left(\frac{a_0}{a}\right)^4,\quad
\rho_{\rm m}(t)=\rho_{m0}\left(\frac{a_0}{a}\right)^3 , \quad
\rho_{\rm h}(t)= \frac{3M_{\rm p}^2\, d^2}{R_{\rm h}^2},
\end{eqnarray}
where the suffix $0$ denotes the value at the present time $t_0$ and $R_{\rm h}$ represents the distance to the future event horizon,
\begin{eqnarray}\label{Rh}
R_{\rm h}(t)\equiv a(t)\int_t^\infty \frac{d a(t')}{H(t')a(t')^2} =
a(t)\int_t^\infty \frac{dt'}{a(t')}.
\end{eqnarray}

The Friedmann equation of the universe with inflaton, radiation, matter, and holographic dark energy is given by
\begin{eqnarray} \label{einstein}
H^2= \frac{\dot a^2}{a^2}
&=&\frac{\rho_{\rm inflaton}+ \rho_{r}+
    \rho_{m}+\rho_{\rm h} }{3M_{\rm p}^2}\,.
\end{eqnarray}
For later convenience, we define the portions of the energy densities at a given time $t$ in the universe by
\begin{eqnarray} \label{eq:Omega}
\Omega_{\rm h} &\equiv& \frac{\rho_{\rm h}}{\rho_{\rm c}}
    =\frac{\rho_{\rm h}}{3 M_{\rm p}^2H^2}
    =\frac{d^2}{H^2 R_{\rm h}^2} ,\\
\Omega_{\rm inflaton}&\equiv&
    \frac{\rho_{\rm inflaton}}{\rho_{\rm c}},
\quad \Omega_{\rm r}\equiv \frac{\rho_{\rm r}}{\rho_{\rm c}},
\quad \Omega_{\rm m}\equiv \frac{\rho_{\rm m}}{\rho_{\rm c}},\nn
\end{eqnarray}
where the critical energy density is $\rho_{\rm c}(t) = 3M_{\rm p}^2 H^2$.
With this definition, the Friedmann equation~(\ref{einstein}) is rewritten as a simple form: $\Omega_{\rm h}+\Omega_{\rm inflaton}+\Omega_{\rm r}+\Omega_{\rm m}=1$.

With the condition $\displaystyle\lim_{t\rightarrow \infty} a(t)=\infty$,  Eq.~(\ref{Rh}) can be casted into the differential form
\begin{eqnarray} \label{dRh:H}
\dot{R}_{\rm h}(t) = H R_{\rm h} -1 \,,
\end{eqnarray}
where the over-dot denotes the derivative with respect to time $t$.
Following the assumption $\dot R_{\rm h}(t)\geq 0$, the event horizon is placed outside of the Hubble radius $[R_{\rm h}(t)\geq H^{-1}(t)]$ always.
From Eqs.~(\ref{dRh:H}) and (\ref{einstein}), a
formula which relates the time derivative of $R_{\rm h}$ with $\Omega_{\rm h}$~\cite{Huang2} can be derived:
\begin{eqnarray} \label{dtRh}
\dot R_{\rm h}(t) &=& \frac{d}{\sqrt{\Omega_{\rm h}}}-1   .
\end{eqnarray}
This equation implies that the distance to the horizon is
a non-decreasing function of time if $d\geq \sqrt{\Omega_{\rm h}(t)}=
\sqrt{1-\Omega_{\rm inflaton}- \Omega_{\rm m}-\Omega_{\rm r}}$ for all $t$.
Comparing Eq.~(\ref{dRh:H}) with Eq.~(\ref{dtRh}), we define the number of e-fold $N_{\rm h}(t)$ of the ratio of the distance to the event horizon and the Hubble radius,
\begin{eqnarray} \label{HRh:dO}
e^{N_{\rm h}(t)}\equiv\frac{R_{\rm h}(t)}{H^{-1}(t)} = \frac{d}{\sqrt{\Omega_{\rm h}}} \geq 1 .
\end{eqnarray}
Note that the second law restricts $d$ to $d\geq 1$ if there is a moment when the holographic dark energy dominates the universe, $\Omega_h=1$.
In this paper, we investigate a possible consequence of the inequality $\dot R_{\rm h}(t)\geq 0$ through the history of our universe.

The Friedmann equation~(\ref{einstein}) is too complex to allow an exact solution.
However, we can develop a good approximation of the evolution by dividing the history of the universe into three pieces as in Fig. 2: the inflation, the consecutive era of RDE and the first half of MDE, and the consecutive era of the last half of MDE and the DDE.
\begin{figure}[tbph]
\begin{center}
\begin{tabular}{ll}
\includegraphics[width=.7\linewidth,origin=tl]{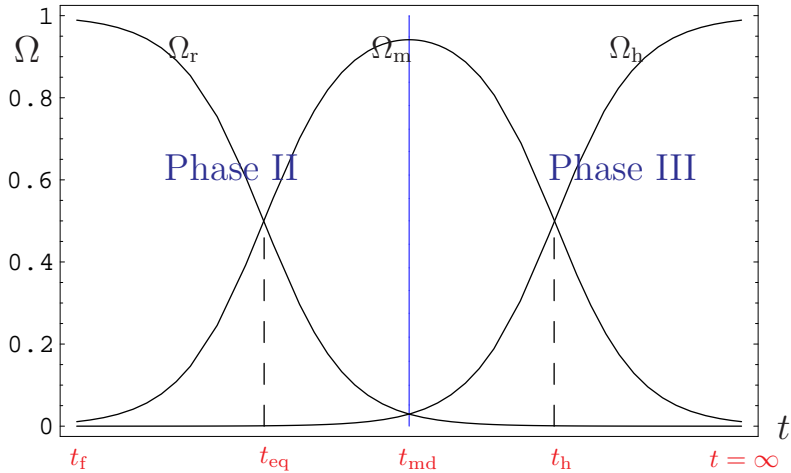}
\end{tabular}
\end{center}
\caption{Schematic diagram for each equipartition times and phases. The phase I is the inflationary era located in the region of times $t< t_{\rm f}$ and is not visualized in this figure. The phases II and III are divided by the equipartition time $t_{\rm md}$ of the radiation and the holographic dark energy.
 } \label{fig2}
\end{figure}

The first phase is the inflation of exponential expansion (phase I) with scale factor
\begin{eqnarray} \label{R:1}
a(t)= a_{\rm i} e^{H_{\rm i} (t-t_{\rm i})}, \quad  t_{\rm i} \le  t < t_{\rm f}-\epsilon,
\end{eqnarray}
where $a_{\rm i}$ is the initial scale factor at $t=t_{\rm i}$, $H_{\rm i}=M_{\rm i}^2/M_P$ is the Hubble parameter with the energy scale $M_{\rm i}$ of the inflation, $\epsilon$ is a short period of time denoting the reheating process after inflation.
The number of e-fold of expansion is $N\equiv H_{\rm i} (t_{\rm f}-t_{\rm i}-\epsilon)$.
During the inflationary phase, we ignore the energy densities of the matter and the radiation compared to the densities of the holographic dark energy and the inflaton energy.
We also assume that there is absent of a time-independent cosmological constant.
Therefore, the energy density of the inflaton field is relevant only during the inflationary period.
At the end of the inflation $(t_{\rm f}-\epsilon \leq t < t_{\rm f})$, there are complex transient phenomena such as preheating and reheating. Rather than dealing with these in detail, we simply assume that the scale factor does not change much during this period.

The phase II ($t_{\rm f}< t\leq t_{\rm md}$) is composed of the consecutive era of RDE and the initial half of MDE, where the subscript ``md" implies the time of matter dominance.
During the phase II, the universe is filled with radiation and matter.
We ignore the densities of the inflaton and the holographic dark energy.
Solving the Friedmann equation~(\ref{einstein}), the scale factor satisfies
\begin{eqnarray} \label{R:2eq}
\left(\y-2 \frac{\rho_{\rm rf}}{\rho_{\rm mf}}\right)
 \sqrt{\y+ \frac{\rho_{\rm rf}}{\rho_{\rm mf}} } = b(t)\equiv\frac{3}{2}\sqrt{\frac{\rho_{\rm mf}}{\rho_{\rm cf}}}H_{\rm f}(t-t_{\rm f})-x_{\rm f}\,,
\end{eqnarray}
where we scaled $\y(t)=\frac{a(t)}{a(t_{\rm f})}$ so that it becomes $1$ at the end of inflation, $H_{\rm f}$,  $\rho_{\rm cf}\equiv 3M_{\rm p}^2 H_{\rm f}^2$, $\rho_{\rm mf}$, and $\rho_{\rm rf}$ are the Hubble parameter, the critical energy density, the matter energy density, and the radiation energy density at time $t_{\rm f}$, respectively.
The integration constant $x_{\rm f}$ is determined by the condition $\y(t_{\rm f})=1$,
\begin{eqnarray}
x_{\rm f} = \left(2 \frac{\rho_{\rm rf}}{\rho_{\rm mf}}-1\right)
 \sqrt{\frac{\rho_{\rm rf}}{\rho_{\rm mf}}+1 } \simeq 2\left(  \frac{\rho_{\rm rf}}{\rho_{\rm mf}}\right)^{3/2} .
\end{eqnarray}
Eq.~(\ref{R:2eq}) allows an explicit exact solution of the scale factor in time:
\begin{eqnarray*} \label{R:2}
\y(t)=\frac1{2^{1/3}}\left[\left(b+\sqrt{b^2-4  \left(\frac{\rho_{\rm rf}}{\rho_{\rm mf}}\right) ^3}\,\right)^{1/3}+
    \left(b-\sqrt{b^2-4 \left(\frac{\rho_{\rm rf}}{\rho_{\rm mf}}\right) ^3}\,\right)^{1/3}\right] .
\end{eqnarray*}

At the initial period of the evolution, the scale factor satisfies $1\leq\y(t)\ll\frac{\rho_{\rm rf}}{\rho_{\rm mf}}$. Then the scale factor becomes
\begin{eqnarray*}
\y(t) &=&\y_0(t)+\frac{\rho_{\rm mf}}{2\rho_{\rm rf}} \left[
    \y_0^2(t)-1 +\frac{1}{3} (\y_0(t)-1)^3 \right]
    +O(\left(\frac{\rho_{\rm mf}}{\rho_{\rm rf}}\right)^{2}),
\end{eqnarray*}
where its zeroth order solution $\y_0(t)$ is the traditional form at RDE:
\begin{eqnarray} \label{y0:rad}
\y_0(t)&=& \left[3\sqrt{\frac{\rho_{\rm rf}}{\rho_{\rm cf}}}H_{\rm f}(t-t_{\rm f}) +1 \right]^{1/2} . \nn
\end{eqnarray}

The transition from RDE to MDE happens when the scale factor satisfies $\y(t_{\rm eq}) = \rho_{\rm rf}/\rho_{\rm mf}$.
After the transition, the scale factor in the limit $\y\gg \rho_{\rm rf}/\rho_{\rm mf}\gg 1$ takes the form of matter dominated era:
\begin{eqnarray} \label{y:matter}
\y(t)&=& \left[\frac32\sqrt{\frac{\rho_{\rm mf}}{\rho_{\rm cf}}}H_{\rm f}(t-t_{\rm f}) -x_{\rm f} \right]^{2/3} +\frac{\rho_{\rm rf}}{\rho_{\rm mf}}+O(t^{-2/3}).
\end{eqnarray}
The phase II ends when the universe is filled with matter with $\Omega_{\rm h}(t_{\rm md}) =\Omega_{\rm r}(t_{\rm md}) \ll 1$.
Therefore, the time of matter dominance, $t_{\rm md}$ is the equipartition time of the radiation and the holographic dark energy.

The phase III ($ t\geq t_{\rm md}$) is the consecutive regime of the matter dominant era and the power-law accelerating era dominated by the dark energy.
The radiation energy density is diluted enough so that it can be ignored relative to the dark energy and the matter in this phase.
We deal with the phase III by using an exact solution as was done by Li in Ref.~\cite{Li}.
Li shows that the function $y=1/\sqrt{\Omega_{\rm h}}$ satisfies the differential equation
\begin{eqnarray} \label{diffeq}
y^2 y'=(1-y^2)\left(\frac{1}{d}+\frac{y}{2}\right),
\end{eqnarray}
where the prime denotes derivative with respect to $\log a$.
He also presented an exact solution to this differential equation for $d=1$.
For arbitrary $d$, the solution of the differential equation~(\ref{diffeq}) was given in Ref.~\cite{Huang},
\begin{eqnarray} \label{sol:3}
\frac{\Omega_{\rm h}(1+\sqrt{\Omega_{\rm h}})^{\frac{d}{2-d}}}{
    (1-\sqrt{\Omega_{\rm h}}
    )^{\frac{d}{d+2}}(d+2\sqrt{\Omega_{\rm h}})^{\frac{8}{4-d^2}}}= x_{\rm md} \,\y(t) .
\end{eqnarray}
The integration constant $x_{\rm md}$ is determined from the junction condition at $t_{\rm md}$.

At the beginning of the phase III, the universe is in MDE and the portion of the holographic dark energy is negligible, $\Omega_{\rm h} \ll 1$. From Eq.~(\ref{sol:3}), it becomes
\begin{eqnarray} \label{Omega:R1}
\Omega_{\rm h}(t)= d^{\frac{8}{4-d^2}}x_{\rm md} \,\y(t) + O(\y^{3/2}).
\end{eqnarray}
We fix the constant $x_{\rm md}$ by comparing the values of $\Omega_{\rm h}(t_{\rm md})$ in phases II and III. Then, we have
\begin{eqnarray}\label{xmd11}
x_{\rm md} \simeq \frac{\Omega_{\rm h}(t_{\rm md})}{ d^{\frac{8}{4-d^2}} \y(t_{\rm md})}
,
\end{eqnarray}
where $\Omega_{\rm h}(t_{\rm md})$ and $\y(t_{\rm md})$ will be specified by the values in phase II.
With Eq.~(\ref{Omega:R1}), the solution of the Friedmann equation~(\ref{einstein}) becomes
\begin{eqnarray*} \label{Rt:Li}
\y(t)=\left[\frac32\sqrt{\frac{\rho_{\rm mf}}{\rho_{\rm cf}}}H_{\rm f}(t-\tau_0)\right]^{2/3}+\frac{2 d^{\frac8{ 4-d^2}}x_{\rm md}}{ 15}
\left[\frac32\sqrt{\frac{\rho_{\rm mf}}{\rho_{\rm cf}}}H_{\rm f}(t-\tau_0)\right]^{4/3}+\cdots\,,
\end{eqnarray*}
where the value of $\tau_0$ is determined from the condition that $\y(t)$ is continuous at $t=t_{\rm md}$.

On the other hand, if the universe is in the DDE, the portion of the holographic dark energy is close to the unity.  In this case, we have $x_{\rm md}\y \gg 1$ and,
from Eq.~(\ref{sol:3}), we get the portion of the holographic dark energy density
\begin{eqnarray} \label{OmegaLambda}
\Omega_{\rm h} = 1-\frac{2^{\frac{4}{2-d}} }{(d+2)^\frac{8}{d(2-d)}
    (x_{\rm md}\y )^\frac{d+2}{d} } +O(\y^{-\frac{2(d+2)}{d}}).
\end{eqnarray}
Now, the Friedmann equation~(\ref{einstein}) becomes, to the present accuracy,
\begin{eqnarray*} \label{eom}
\frac{2^{\frac{4}{2-d}}H^2 }{(d+2)^\frac{8}{d(2-d)}
    ( x_{\rm md}\y)^\frac{d+2}{d} }
= \frac{\rho_{\rm mf}}{\rho_{\rm cf}}
        \frac{H_{\rm f}^2}{\y^3} .
\end{eqnarray*}
The solution to this equation is
\begin{eqnarray} \label{sol:Li}
x_{\rm md}\y(t) =  \left(\frac{(d-1) S}{d}\right)^{\frac{d}{d-1}}
    \left[H_f(t-t_D)\right]^{\frac{d}{d-1}},
\end{eqnarray}
where $t_D$ is a constant of smaller scale than the typical value of $t$ in the DDE and will be specified from the next order calculation and
\begin{eqnarray*} \label{RD}
 S= \frac{(d+2)^{\frac4{d(2-d)}}}{2^{\frac{2}{2-d}}}\,\sqrt{\frac{\rho_{\rm mf}}{\rho_{\rm cf}}}\, x_{\rm md}^{3/2}.
\end{eqnarray*}
Since $x_{\rm md}$ is given in Eq.~(\ref{xmd11}), Eq.~(\ref{sol:Li}) determines the asymptotic evolution of $R(t)$ almost completely.
Although it is well known that the scale factor is proportional to $t^{d/(d-1)}$, the proportionality constant $S$ is determined here for the first time.

By dividing the evolution of the universe into the three phases we get the evolution of the scale factor in a very accurate form.
Especially, the energy densities of the neglected components are extremely small for each phases. Explicitly for the phase III $(t>t_{\rm md})$, we show in the next section that the maximum value of the portion of the neglected radiation is given by $\Omega_{\rm r}(t_{\rm md}) \sim \frac{\rho_{\rm rf}}{\rho_{\rm mf}}\,e^{-N_{\rm h}(t_{\rm f})} $ at time $t=t_{\rm md}$. Therefore, the relative error ($< 10^{-3}$) of the present approximation is very small. This is why we deal the Friedmann equation rather with this somewhat complex form than with taking a simpler approximation.

\section{Cosmological history} \label{sec:3}
In the previous section, we have obtained the evolution of the scale factor of the universe with the holographic dark energy by dividing the evolution into three pieces and solving the Friedmann equation.
In this section, we calculate the detailed evolution of physical parameters such as the energy densities, the distance to the future event horizon, and the Hubble parameter, for each phase.

\subsection{The inflationary phase}\label{sec:3.1}
As usual, we assume that the Hubble parameter in this phase is nearly constant,
\begin{eqnarray*}
H(t)= H_{\rm i},\quad\quad 
    t < t_{\rm f}-\epsilon.
\end{eqnarray*}
At the beginning of the inflation, there may present some portions of matters, radiation, inflaton, and the holographic dark energy. During the exponential expansion of the scale factor, the densities of the holographic dark energy and the inflaton change slowly.
However, the densities of the matter and the radiation decrease exponentially with time and at some time $t_{\rm i}$ they become effectively negligible.
We discuss the inflation starting from this time $t_{\rm i}$.
For $t\geq t_{\rm i}$, the portions of the energies satisfy,
\begin{eqnarray*} \label{rho:inflation}
\Omega_{\rm hi}+\Omega_{\rm inflaton}(t_{\rm i})
= 1= \Omega_{\rm h}(t_{\rm f}-\epsilon)+\Omega_{\rm inflaton}(t_{\rm f}-\epsilon) ,
\end{eqnarray*}
where $\Omega_{\rm hi}\equiv\Omega_{\rm h}(t_{\rm i})$ is the initial portion of the holographic dark energy.
The portion of the inflaton energy at the end of inflation is
\begin{eqnarray} \label{rhoInflaton}
\Omega_{\rm inflaton}(t_{\rm f}-\epsilon)=
    \frac{\rho_{\rm inflaton}(t_{\rm f}-\epsilon)}{3M_{\rm p}^2H_{\rm i}^2}
 = 1-\Omega_{\rm h}(t_{\rm f}-\epsilon).
\end{eqnarray}

From the scale factor~(\ref{R:1}) and the definition~(\ref{Rh}), the distance to the cosmological event horizon during the phase I is
\begin{eqnarray} \label{Rhsimple}
R_{\rm h}(t)= \frac{1}{H_{\rm i}}\left(1+ C\,e^{H_{\rm i}(t-t_{\rm i})}\right).
\end{eqnarray}
The parameter $C$ is an integration constant to be determined from the initial portion of the holographic energy at time $t_{\rm i}$,
\begin{eqnarray*} \label{A}
 C= H_{\rm i} R_{\rm h}(t_{\rm i})-1= \dot{R}_{\rm h}(t_{\rm i})
  =\frac{d}{\sqrt{\Omega_{\rm hi}}}-1 \geq0 \,,
\end{eqnarray*}
where the second and third equalities come from Eqs.~(\ref{dRh:H}) and (\ref{dtRh}), respectively.
At time $t_{\rm i}$, we have no criteria to specify the value of the portion of the holographic dark energy.
Since the distance to the horizon grows as time as in Eq.~(\ref{Rhsimple}), the holographic dark energy will gradually be transferred to the inflaton energy to satisfy the equalities in Eq.~(\ref{rho:inflation}).

How can we determine the initial distance to the horizon $R_{\rm h}(t_{\rm i})$?
It may be determined by measuring the initial energy densities of the holographic dark energy and the inflaton field.
Since $\Omega_{\rm h}(t)$ should be not larger than one, we have  constraint for  $C\geq d-1$. This restricts the value of $d$ into
\begin{eqnarray*}
\sqrt{\Omega_{\rm hi}} \leq d \leq C+1.\nn
\end{eqnarray*}
 If initially the inflaton field is in its vacuum state so that $\Omega_{\rm hi}=1$, we may have the identity $ C=d-1$.
For $d=1$ in this case, we may have $C=0$, which implies a permanent exponential inflation.

In the presence of the inflaton field, the horizon may not be kept at a constant distance but goes away. Therefore, it is natural to assume that the value of $C$ is positive and the distance to the future event horizon grows during the inflation.
At $t=t_{\rm f}-\epsilon$,
\begin{eqnarray} \label{Rh:f}
R_{\rm h}(t_{\rm f}-\epsilon)
  &=&\frac{1}{H_{\rm i}} \left[1+ \left(\frac{d}{\sqrt{\Omega_{\rm hi}}}-1\right)e^N\right] = \frac{e^{N_{\rm h}(t_{\rm f}-\epsilon)}}{H_{\rm i}}  \,,\\
\dot R_{\rm h}(t_{\rm f}-\epsilon)
  &=& \left(\frac{d}{\sqrt{\Omega_{\rm hi}}}-1\right)e^N\simeq e^{N_{\rm h}(t_{\rm f}-\epsilon)}, \nn
\end{eqnarray}
where the the number of e-fold~(\ref{HRh:dO}) at time $t_{\rm f}-\epsilon$ is
\begin{eqnarray}
N_{\rm h}(t_{\rm f}-\epsilon)= N+\log\left(\frac{d}{\sqrt{\Omega_{\rm hi}}}-1+e^{-N} \right).
\end{eqnarray}
Note that Eq.~(\ref{Rh:f}) restricts the value of $d$ to $d\geq \sqrt{\Omega_{\rm hi}}$ if we want the distance to the horizon to be positive after the sufficient inflation.

We also assume that a sufficient expansion of the horizon happens, $e^{N_{\rm h}}\gg 1$.
In the presence of a fine tuning of $d$ close to $\sqrt{\Omega_{\rm hi}}$,  it would be possible that the distance to the event horizon does not vary much compared to the change of the scale factor. However, we ignore this possibility.
Physically, this is correct since the density of the holographic dark energy is negligible just after the inflation.
The density of the inflaton at the end of the inflation becomes
\begin{eqnarray} \label{rho:inflatontf}
\rho_{\rm inflaton}(t_{\rm f}-\epsilon)
    = \rho_{\rm ci}[1-\Omega_{\rm h}(t_{\rm f}-\epsilon)]
    = \rho_{\rm ci} -\frac{d^2\rho_{\rm ci}}{e^{2N_{\rm h}(t_{\rm f}-\epsilon)}} .
\end{eqnarray}
where $\rho_{\rm ci} = \rho_{\rm c}(t_{\rm i})=3M_p^2 H_i^2$. In the presence of a sufficient expansion of the scale factor, we ignore the second term of the right hand side of Eq.~(\ref{rho:inflatontf}) and get $\rho_{\rm inflaton}(t_{\rm f}-\epsilon)\simeq \rho_{\rm ci}$.

In reality, one should solve both the inflaton field equation and the Friedmann equation to describe the inflationary period precisely.
However, in this paper, we simply assume that there is an exponential inflation.
In Ref.~\cite{Chen}, it was shown that there is an inflationary solution very close to this assumption with a single minimally coupled inflaton field.

At the end of an inflation, there happen many things such as preheating and reheating.
We assume that the processes happen during the period $t_{\rm f}-\epsilon < t < t_{\rm f}$.
We do not deal with these processes in detail and simply present the values of the energy densities after the process.
In fact, the quantities of the energy densities after the inflation depend on the detailed process of reheating.
If the reheating process happens almost instantaneously ($\epsilon\to 0$; instantaneous reheating approximation),  the total energy density does not vary much during the process. Therefore, the physical parameters such as the total energy density, the Hubble parameter, and $ \dot R_{\rm h}$ must be continuous at $t_{\rm f}$.

On the other hand, if the period lasts somewhat longer period of time (usually the universe during the reheating is assumed to be filled with the matter of inflaton oscillation), the Hubble parameters  and the distance to the horizon at times $t_{\rm f}-\epsilon$ and $t_{\rm f}$ are different from each other.
The changes of these physical parameters should be taken into account in this case.
Rather than calculating the changes, we write the resulting scale factor to be  $R(t_{\rm f})$ and assume that the densities of the holographic dark energy, the matter, and the radiation satisfy
\begin{eqnarray}\label{assum1}
\rho_{\rm rf} \gg \rho_{\rm mf} \gg \rho_{\rm hf},
\end{eqnarray}
where $\rho_{\rm hf}=\rho_{\rm h}(t_{\rm f}) $ is the density of the holographic dark energy at time $t_{\rm f}$.
The holographic dark energy at $t_{\rm f}-\epsilon$ is  exponentially small because of the exponential factor in $R_{\rm h}$.
Therefore, the number of e-fold $N_{\rm h}(t_{\rm f})$ in Eq.~(\ref{HRh:dO}) becomes
\begin{eqnarray} \label{Rh:f2}
N_{\rm h}(t_{\rm f}) &=& \log (H_f \,R_{\rm h}(t_{\rm f}))\\
&=&N_{\rm h}(t_{\rm f}-\epsilon)+\log\left(\frac{H_{\rm f}\, R_{\rm h}(t_{\rm f})}{H_{\rm i}R_{\rm h}(t_{\rm f}-\epsilon)}\right)  \nn
.
\end{eqnarray}
Since the change of the e-fold of the scale factor during the reheating phase is much smaller than that of the inflationary phase, we may also use the condition for sufficient expansion of $R_{\rm h}$ so that $e^{N_{\rm h}(t_{\rm f})}\gg 1$.
Therefore, the portion of the holographic dark energy at time $t_{\rm f}$
\begin{eqnarray} \label{Omegah:f}
\Omega_{\rm h}(t_{\rm f})=\frac{d^2}{e^{2N_{\rm h}(t_{\rm f})} }\ll 1
\end{eqnarray}
is extremely small.

\subsection{From RDE to MDE}
    \label{sec:3.2}
Now consider the phase II ($t_{\rm f}\leq t\leq t_{\rm md}$) which is composed of the whole RDE and the first half of MDE.
At the end of the inflationary phase, most of the inflaton energy have changed to the radiation.
During this phase, we ignore the holographic dark energy when we calculate the evolution of $\y(t)$.
We calculate the evolution of the distance to the horizon simply as if we are in a  Robertson-Walker universe with radiation and matter.
The portion of the holographic dark energy density gradually increases and will be maximized at the end of this phase, $t=t_{\rm md}$. However, the maximal value is of the order of $\Omega_{\rm mf}^{-1}e^{-N_{\rm h}(t_{\rm f})}$ justifying the present approximation to the accuracy.

The Hubble parameter in this phase is
\begin{eqnarray} \label{H:2}
H(t)=\frac{\dot{\y}}{\y}= H_{\rm f}\sqrt{\frac{\rho_{\rm mf}}{\rho_{\rm cf}}}
    \,\frac{\sqrt{\y+ \rho_{\rm rf}/\rho_{\rm mf}}}{\y^2}\,,
    \quad t_{\rm f}< t\leq t_{\rm md}.
\end{eqnarray}
At the beginning of this phase $t=t_{\rm f}$, the Hubble parameter becomes
\begin{eqnarray} \label{dR:tf+}
H(t_{\rm f}) = H_{\rm f},
\end{eqnarray}
where we use $\rho_{\rm rf}+\rho_{\rm mf}=\rho_{\rm cf}$.
The holographic dark energy density is ignored in this equation.

In the instantaneous reheating approximation, both of the Hubble parameter and $ \dot R_{\rm h}$ must be continuous at $t_{\rm f}$ and we obtain the total energy density at the beginning of RDE from the initial conditions:
\begin{eqnarray} \label{rho:rhoi}
\rho_{\rm rf}+\rho_{\rm mf} = \rho_{\rm inflaton}(t_{\rm f}-\epsilon)
\simeq \rho_{\rm ci}-\frac{d^2\rho_{\rm ci}}{e^{2N_{\rm h}(t_{\rm f})}} .
\end{eqnarray}
For large $N_{\rm h}(t_{\rm f})$, we have $\rho_{\rm rf}+\rho_{\rm mf} \simeq \rho_{\rm ci}$.
On the other hand, if the period lasts for a non-negligible period of time, we use the assumption in Eq.~(\ref{assum1}), which leads to the inequality
\begin{eqnarray} \label{cond}
e^{2N_{\rm h}(t_{\rm f})}\gg \frac1\Of \gg 1 ,
\end{eqnarray}
where $\Omega_{\rm mf}\equiv \frac{\rho_{\rm mf}}{\rho_{\rm rf}+\rho_{\rm mf}}\simeq\frac{\rho_{\rm mf}}{\rho_{\rm rf}}\simeq \frac{\rho_{\rm mf}}{\rho_{\rm cf}} $ denotes the portion of the matter energy at the time $t_{\rm f}$ and $\rho_{\rm cf} = 3M_{\rm p}^2 H^2(t_{\rm f})$.
From this point on in this paper, we assume that $\rho_{\rm rf}\simeq\rho_{\rm cf}$ for simplicity and the stronger constraint
\begin{eqnarray}\label{cond:str}
e^{N_{\rm h}(t_{\rm f})}\gg \frac{d}{\Of}
\end{eqnarray}
is satisfied with the parameters $N_{\rm h}(t_{\rm f})$, $d$, and  $\Omega_{\rm mf}$.
In fact, in the next subsection, it turns out that Eq.~(\ref{cond:str}) guarantees the presence of  MDE between RDE and DDE.
If we have explicit model of inflation and reheating, we may determine $H_{\rm f}$ and $t_{\rm f}$ from the initial conditions.

The densities of the radiation and the matter decrease as $1/a^4$ and $1/a^3$, respectively. Therefore, the densities at time $t$ become
\begin{eqnarray} \label{rhoR:t}
\rho_{\rm r}(t)&=& \frac{\rho_{\rm rf}}{\y^4(t)}, \quad
\rho_{\rm m}(t)= \frac{\rho_{\rm mf}}{\y^3(t)} .
\end{eqnarray}
The transition to MDE happens at time $t_{\rm eq}$ when $\rho_{\rm r}(t_{\rm eq})=\rho_{\rm m}(t_{\rm eq})$. The scale factor at this time is
\begin{eqnarray}
\y(t_{\rm eq})= \frac1{\Of}.
\end{eqnarray}

Using $a(t)$ in Eqs.~(\ref{Rh}) and (\ref{H:2}), we obtain the distance to the horizon from Eq.~(\ref{Rh}):
\begin{eqnarray} \label{Rh2}
R_{\rm h}(t) &=&\y\left(\int_1^\infty\frac{d\y}{H \y^2}
    -\int_{1}^\y\frac{d\y}{H\y^2}\right) \\
&= & \frac{1}{H_{\rm f}\Omega_{\rm mf}} \left(g_N-2
    \sqrt{\Of\y+ 1}\right)\y\nn,
\end{eqnarray}
where we use $\int_1^\infty\frac{d\y}{H \y^2}=R_{\rm h}(t_{\rm f}) $ and $g_N$ is given by
\begin{eqnarray}\label{gN:N}
g_N &=&
   \Of\,e^{N_{\rm h}(t_{\rm f})}+ 2\sqrt{1+\Of}\simeq \Of e^{N_{\rm h}(t_{\rm f})}.
\end{eqnarray}
In the second equality, we use Eq.~(\ref{cond:str}).

In the phase II, we cannot use the formula~(\ref{dtRh}) because we have ignored the holographic dark energy to get the solution of the Friedmann equation.
Instead, the density of the holographic dark energy is given by scaling $R_{\rm h}(t)$ and it becomes
\begin{eqnarray} \label{rhoh:2}
\rho_{\rm h}(t)=\rho_{\rm h}(\tf) \frac{R_{\rm h}^2(t_{\rm f})}{R_{\rm h}^2(t)}=
   \frac{\rho_{\rm cf}d^2\Of^2}{\y^2\left(g_N-
    2\sqrt{\Of \y+1}\right)^{2}} .
\end{eqnarray}
The phase II ends at time $t_{\rm md}$ when $\rho_{\rm h}(t_{\rm md}) = \rho_{\rm r}(t_{\rm md})$.  The scale factor at this time is
\begin{eqnarray} \label{y:teq}
\y(t_{\rm md})
    &=&\frac{g_N+2/d-\sqrt{d^{-1}g_N+1+ d^{-2}}}{d\,\Of} \simeq \frac{e^{N_{\rm h}(t_{\rm f})}}{d}.
\end{eqnarray}
Interestingly, the scale factor $\y(t_{\rm md})$ is dependent on $N_{\rm h}(t_{\rm f})$ rather than $N$.
In addition, it is almost independent of the other physical parameters such as $\Omega_{\rm mf}$.
Using the approximate formula for $\y(t)$ in Eqs.~(\ref{y:matter}), and (\ref{y:teq}) we get the time of full matter dominance,
\begin{eqnarray} \label{teq:AN}
 t_{\rm md}\simeq  \frac{2\,e^{\frac{3N_{\rm h}(t_{\rm f}) }2} }{3 H_{\rm f}d^{3/2}\Of^{1/2}}.
\end{eqnarray}
Note that this time is dependent on $N_{\rm h}(t_{\rm f})$ rather than $N$ itself.

The Hubble parameter~(\ref{H:2}) at this time is
\begin{eqnarray} \label{H:teq}
H(t_{\rm md})= H_{\rm f}
    \sqrt{\frac{\rho_{\rm mf}}{\rho_{\rm cf}}} \frac{\sqrt{\y(t_{\rm md})+ \Of^{-1}}}{\y^2(t_{\rm md})}
\simeq H_{\rm f} \frac{d^{3/2}\sqrt{\Of}}{e^{\frac{3N_{\rm h}(t_{\rm f})}2}} .
\end{eqnarray}
The energy densities at $t_{\rm md}$ becomes
\begin{eqnarray} \label{rhos}
\rho_{\rm r}(t_{\rm md})&=&\rho_{\rm h}(t_{\rm md})\simeq
    \rho_{\rm cf} d^4\,e^{-4N_{\rm h}(t_{\rm f})} \,, \\
\rho_{\rm m}(t_{\rm md})
  &\simeq&\rho_{\rm cf}d^3\,\Of \, e^{-3N_{\rm h}(t_{\rm f})}. \nn
\end{eqnarray}
At the time of matter dominance $t_{\rm md}$, we should have $\rho_{\rm r}(t_{\rm md}) \ll \rho_{\rm m}(t_{\rm md})$.
This provides the condition~(\ref{cond:str}).
The portions of the holographic dark energy, the radiation, and the matter at time $t_{\rm md}$ are
\begin{eqnarray} \label{Omega:teq}
\Omega_{\rm h}(t_{\rm md})=\Omega_{\rm r}(t_{\rm md}) &=& \frac{\rho_{\rm r}(t_{\rm md})}{3M_p^2 H^2(t_{\rm md})} \simeq \frac{d}{\Of e^{N_{\rm h}(t_{\rm f})}}\ll 1,\\
\Omega_{\rm m}(t_{\rm md})& \simeq&  1-\frac{d}{\Of  e^{N_{\rm h}(t_{\rm f})}}.
\nn
\end{eqnarray}

The time derivative of $R_{\rm h}$ is
\begin{eqnarray} \label{dRh2}
\dot{R}_{\rm h}(t)=
    \frac{g_N\sqrt{\Of\y+1}}{\Of\y}-3 -\frac{ 2}{\Of\y} \,.
\end{eqnarray}
The second law of thermodynamics says that the value of $\dot{R}_{\rm h}(t)$ should be non-negative.
However, Eq.~(\ref{dRh2}) becomes negative since $\y$ indefinitely increases with time.
The time derivative $\dot{R}_{\rm h}(t)$ vanishes at time $t_{\rm max}$ where the scale factor becomes
\begin{eqnarray} \label{tmax}
\y(t_{\rm max})&=& \frac{g_N^2+6+ g_N\sqrt{g_N^2+12 }}{18\Of}
    \simeq\frac{1}{9} \Of\,e^{2N_{\rm h}(t_{\rm f})} .
\end{eqnarray}
If the universe is still in MDE after the time $t_{\rm max}$,  the distance to the horizon decreases for $t>t_{\rm max}$.
This faulty behavior is due to the failure of the present approximation scheme ignoring the holographic dark energy in the phase II.
Therefore,  the holographic dark energy should be included before the time $t_{\rm max}$ to have accurate solution of the Friedmann equation.
Naturally, the phase II should not include this time region and we have the restriction $\y(t_{\rm md}) \ll \y(t_{\rm max})$, which is respected by the condition~(\ref{cond:str}).
As seen in the present calculation, the decelerating expansion makes $\dot R_{\rm h}$ decrease and the accelerating expansion increase.
This conclusion is in contrast with the apparent facts:  In the permanent exponential expansion of de-Sitter space, the distance to the cosmological event horizon $R_{\rm h}$ is a constant of time. In decelerating power law expansion, the event horizon is present at infinity.

\subsection{From Matter Dominant Era to Dark Energy Dominant Era} \label{sec:3.3}
At the beginning of the phase III, the portion of the radiation energy is already negligible ($\Omega_{\rm r}(t_{\rm md}) \sim d \Omega_{\rm mf}^{-1} \,e^{-N_{\rm h}(t_{\rm f})}$) and keeps decreasing throughout the whole evolution. The matter energy density dominates the first stage evolution.
The portion of the holographic dark energy is negligible at the beginning, however, it keeps increasing throughout the whole evolution of the phase III.
It is the same as that of the portion of the radiation at $t_{\rm md}$ and becomes the same as that of the matter at $t_{\rm h}(> t_{\rm md})$.
In this sense, we ignore the radiation in phase III compared to the dark energy and the matter.

The portion of the holographic dark energy satisfies $\Omega_{\rm h}(t_{\rm md})\ll 1$ since we are in MDE at this time.
Since $\Omega_{\rm h}(t)$ is continuous at $t_{\rm md}$, from Eqs.~(\ref{Omega:R1}), (\ref{Omega:teq}), and (\ref{y:teq}), we have
\begin{eqnarray} \label{xmd}
\y(t_{\rm md})\simeq  \frac{e^{N_{\rm h}(t_{\rm f})}}{d},
\quad x_{\rm md} =
    \frac{1}{\Of} \frac{d^{-\frac{2d^2}{4-d^2}}}{e^{2N_{\rm h}(t_{\rm f})}} .
\end{eqnarray}

In this phase, it would be better to use $\Omega_{\rm h}$ instead of $\y$ as a parameter characterizing a given moment of time.
From Eqs.~(\ref{sol:3}) and (\ref{xmd}), the relative scale factor $\y$ can be rewritten in terms of  $\Omega_{\rm h}$ as,
\begin{eqnarray}\label{a:Omega}
\y(\Omega_{\rm h}) = \frac{\Omega_{\rm mf} e^{2N_{\rm h}(t_{\rm f})}}{d^{2}}
      \frac{\Omega_{\rm h}(1+\sqrt{\Omega_{\rm h}})^{\frac{d}{2-d}}}{
    (1-\sqrt{\Omega_{\rm h}}
    )^{\frac{d}{d+2}}(1+2\sqrt{\Omega_{\rm h}}/d)^{\frac{8}{4-d^2}}} .
\end{eqnarray}
From the Friedmann equation, the Hubble parameter becomes
\begin{eqnarray} \label{H:Li}
H = \frac{H_{\rm f}}{\Omega_{\rm mf}e^{3N_{\rm h}(t_{\rm f})}}
     \left[ d^{2}\,
 \frac{(1-\sqrt{\Omega_{\rm h}}
    )^{\frac{d}{d+2}}(1+2\sqrt{\Omega_{\rm h}}/d)^{\frac{8}{4-d^2}}}
 {\Omega_{\rm h}(1-\Omega_{\rm h})^{1/3}(1+\sqrt{\Omega_{\rm h}})^{\frac{d}{2-d}}}
 \right]^{\frac32} .
\end{eqnarray}
The explicit value of $d$ should be taken to be larger than one since there is no inflaton field.

At the final stage of the evolution, with $\y\rightarrow \infty$ and $d\neq 1$,  from Eq.~(\ref{sol:Li}), we have
\begin{eqnarray}\label{HRh:final}
H(t) = \frac{d}{(d-1)(t-t_D)}, \quad R_{\rm h}(t) = (d-1) (t-t_D) .
\end{eqnarray}
Therefore, the value of $d$ governs the final evolution.

Now let us consider the physics at the equipartition time $t_{\rm h}$ of the matter and the holographic dark energy.
The holographic dark energy is treated exactly in the phase III.
Therefore, the time derivative $\dot R_{\rm h}$ is related to the holographic dark energy density through Eq.~(\ref{dtRh}).
Since we ignore the radiation energy density, we have $\Omega_{\rm h}+\Omega_{\rm m}=1$.
At the time $t_{\rm h}$, the portions of the holographic dark energy and the matter are the same: $\Omega_{\rm h}(t_{\rm h})=1/2=\Omega_{\rm m}(t_{\rm h})$.
After this time ($t> t_{\rm h}$), the holographic dark energy starts to dominate the unverse.
The relative scale factor at this time is given by
\begin{eqnarray} \label{Rth}
\y(t_{\rm h}) &=&
    \Of \,e^{2N_{\rm h}(t_{\rm f}) }\,\frac{ c(d)}{d^2},
\end{eqnarray}
where $c(d)$ is a non-decreasing function of $d\geq 0$ only:
\begin{eqnarray}
c(d)&=&
\left(\frac{(1+\sqrt{2})^{d}}{2^{d^2/4} (1+\sqrt{2}/d)^{2}}\right)^{\frac{4}{4-d^2}} , \nn
\end{eqnarray}
which varies from $0$ to $2$ as $d$ changes from $0$ to $\infty$.
The Hubble parameter and the distance to the horizon at $t_{\rm h}$ are given by
\begin{eqnarray} \label{Hubble:eq}
H(t_{\rm h}) = \frac{\sqrt{2} H_{\rm f}}{\Of } \frac{d^3}{c^{3/2}(d) \,e^{3N_{\rm h}(t_{\rm f})}}, \quad
R_{\rm h}(t_{\rm h}) = \frac{\Of}{H_{\rm f}}\frac{c^{3/2}(d) \,e^{3N_{\rm h}(t_{\rm f})}}{d^2}.
\end{eqnarray}
The ratio of the distance to the event horizon to the Hubble radius is $H(t_{\rm h}) R_{\rm h}(t_{\rm h})=\sqrt{2} d$.

On the other hand, one may calculate when does the acceleration of the scale factor $a(t)$ become positive by calculating the deceleration parameter.
Instead of direct calculation, to get $\ddot a= a(\dot H+H^2)$ we use
\begin{eqnarray} \label{dOmega}
&&-2 \left(\frac{\dot H}{H^2} +1-\frac{\sqrt{\Omega_{\rm h}}}{d} \right)=\frac{\dot\Omega_{\rm h}}{H\Omega_{\rm h}}
    = \frac{1}{g(d,\Omega_{\rm h})}; \\
&&g(d,\Omega_{\rm h})=1+\frac{d}{2(2-d)}
    \frac{\sqrt{\Omega_{\rm h}}}{1+ \sqrt{\Omega_{\rm h}}}
    +\frac{d}{2(d+2)}\frac{\sqrt{\Omega_{\rm h}}}{1- \sqrt{\Omega_{\rm h}}}
    -\frac{8}{4-d^2}\frac{\sqrt{\Omega_{\rm h}}}{d+ 2\sqrt{\Omega_{\rm h}}} \nn ,
\end{eqnarray}
where the first equality of the first line of Eq.~(\ref{dOmega}) comes from the definition of holographic dark energy density~(\ref{eq:Omega}) and the second equality from Eq.~(\ref{sol:3}).
The deceleration parameter now becomes
\begin{eqnarray} \label{ddR}
q(t) \equiv-\frac{\ddot{a}}{aH^2}=-\frac{\sqrt{\Omega_{\rm h}}}{d}+
    \frac{1}{2g(d,\Omega_{\rm h})} .
\end{eqnarray}
Even though the equation $\ddot{a}(t)=0$ allows a closed form of solution,
we write an approximate solution $\Omega_{\rm t}\equiv\Omega_{\rm h}\simeq 0.432+0.145(d-1)$ around $d\sim 1$.
This result implies that the value of $\y(t_{\rm h})$ will be of the same order as $\y(\Omega_{\rm t})$.

The radiation energy density in the phase III is
\begin{eqnarray}
\rho_{\rm r} (t) &=& \frac{\rho_{\rm cf}}{\Omega_{\rm mf}^4 e^{8N_{\rm h}(t_{\rm f})}} \left[d^{2}\,
 \frac{(1-\sqrt{\Omega_{\rm h}}
    )^{\frac{d}{d+2}}(1+2\sqrt{\Omega_{\rm h}}/d)^{\frac{8}{4-d^2}}}
 {\Omega_{\rm h}(1+\sqrt{\Omega_{\rm h}})^{\frac{d}{2-d}}}
 \right]^4\,.
\end{eqnarray}
The matter energy density is
\begin{eqnarray}
\rho_{\rm m} (t) &=& \frac{\rho_{\rm cf}}{\Omega_{\rm mf}^2 e^{6N_{\rm h}(t_{\rm f})}}  \left[ d^{2}\,
 \frac{  (1-\sqrt{\Omega_{\rm h}}
    )^{\frac{d}{d+2}}(1+2\sqrt{\Omega_{\rm h}}/d)^{\frac{8}{4-d^2}}}
 {\Omega_{\rm h}(1+\sqrt{\Omega_{\rm h}})^{\frac{d}{2-d}}}
 \right]^3\,.
 \end{eqnarray}
The density of the holographic dark energy is
\begin{eqnarray}
\rho_{\rm h}(t) &=&
    \frac{\rho_{\rm cf}}{\Omega_{\rm mf}^2 e^{6N_{\rm h}(t_{\rm f})}}
     \left[ d^{2}\,
 \frac{(1-\sqrt{\Omega_{\rm h}}
    )^{\frac{d}{d+2}}(1+2\sqrt{\Omega_{\rm h}}/d)^{\frac{8}{4-d^2}}}
 {\Omega_{\rm h}^{2/3}(1-\Omega_{\rm h})^{1/3}(1+\sqrt{\Omega_{\rm h}})^{\frac{d}{2-d}}}
 \right]^{3} .
\end{eqnarray}
The ratio of the holographic dark energy and the matter energy density
$\rho_{\rm h}(t)/\rho_{\rm m}(t) = \Omega_{\rm h}(t)/(1-\Omega_{\rm h}(t))$ is independent of $d$ and well met with the criteria $\Omega_{\rm h}+\Omega_{\rm m}=1$.

We may determine the parameters $\Omega_{\rm mf}$, $N_{\rm h}(t_{\rm f})$, and $\rho_{\rm cf}$ from the present data of the universe.
Let us set the present values of the holographic dark energy, the ratio of the densities of the radiation and matter, the Hubble parameter, the deceleration parameter, and the relative scale factor to be $\Omega_{\rm h}(t_0)\equiv \Omega_{\rm h0}$,  $\Omega_{\rm r}(t_0)/\Omega_{\rm m}(t_0) =r_0$, $H(t_0)= H_0$, $q(t_0)=q_0$, and $\y(t_0)=\y_0$.
Note that the deceleration parameter~(\ref{ddR}) is dependent on $d$ and $\Omega_{\rm h}$ and independent on the other energy densities. Therefore, once we measure the deceleration parameter and the portion of the holographic dark energy at the present time, we may get the explicit value of $d$ from
\begin{eqnarray} \label{ddR}
q(t_0) =-\frac{\sqrt{\Omega_{\rm h0}}}{d}+
    \frac{1}{2g(d,\Omega_{\rm h0})} .
\end{eqnarray}
If we use the present data $q(t_0)\sim -1$ and $\Omega_{\rm h0}\sim 0.72$, we get $d\sim 0.6$. On the other hand, if we want to have $d\sim 1$, we need $q(t_0)\sim -0.5$. This value is within $1\sigma$ error of the present experimental data ($-0.96\pm 0.43$)~\cite{wang}.

From the ratio $r_0=\rho_{\rm r}(t_0)/\rho_{\rm m}(t_0)$, we have
\begin{eqnarray}\label{OeN}
\Omega_{\rm mf} e^{N_{\rm h}(t_{\rm f})} &=&
    \frac1{\sqrt{r_0}}\left[d^{2}\,
 \frac{(1-\sqrt{\Omega_{\rm h0}}
    )^{\frac{d}{d+2}}(1+2\sqrt{\Omega_{\rm h0}}/d)^{\frac{8}{4-d^2}}}
 {\Omega_{\rm h0}(1+\sqrt{\Omega_{\rm h0}})^{\frac{d}{2-d}}}
 \right]^{1/2} .
\end{eqnarray}
 Using Eqs.~(\ref{a:Omega}) and (\ref{OeN}), we may identify the e-fold of the ratio between the distance to the horizon and the Hubble radius,
\begin{eqnarray}
e^{N_{\rm h}(t_{\rm f}) }=
    \sqrt{r_0} \y_0 \left[d^{2}\,
 \frac{(1-\sqrt{\Omega_{\rm h0}}
    )^{\frac{d}{d+2}}(1+2\sqrt{\Omega_{\rm h0}}/d)^{\frac{8}{4-d^2}}}
 {\Omega_{\rm h0}(1+\sqrt{\Omega_{\rm h0}})^{\frac{d}{2-d}}}
 \right]^{3/2} .
\end{eqnarray}
 From Eqs.~(\ref{H:Li}) and (\ref{OeN}),  we get
\begin{eqnarray}
H_{\rm f}&=&
    H_0\sqrt{r_0} \y_0^2\sqrt{1-\Omega_{\rm h0}}\,
    \left[d^{2}\,
 \frac{(1-\sqrt{\Omega_{\rm h0}}
    )^{\frac{d}{d+2}}(1+2\sqrt{\Omega_{\rm h0}}/d)^{\frac{8}{4-d^2}}}
 {\Omega_{\rm h0}(1+\sqrt{\Omega_{\rm h0}})^{\frac{d}{2-d}}}
 \right]^{2} .
\end{eqnarray}
In this way, we may identify all initial parameters at time $t_{\rm f}$ from the data today.

Using the present data, $\Omega_{\rm h0}\simeq 0.72$ and $r_0 \simeq 10^{-4}$, we may get more explicit value. For example, Eq.~(\ref{OeN}) becomes
\begin{eqnarray}\label{OeN2}
\Omega_{\rm mf} e^{N_{\rm h}(t_{\rm f})}\simeq
 138.9\cdot d(0.389)^{\frac{d}{d+2}}\left(\frac{1+1.697/d}
 {(1.8485)^{\frac{d(d+2)}{8}}}\right)^{\frac{4}{4-d^2}}. \nn
\end{eqnarray}
The function  $ \Omega_{\rm mf} e^{N_{\rm h}(t_{\rm f})}$ is approximated by $280.0+61.2(d-1)$ around $d=1$.
This shows that the stronger constraint~(\ref{cond:str}) is valid for our universe.

\section{Summary and Discussions} \label{sec:4}
The precise history of the universe  is presented in the presence of the inflaton, the matter, the radiation, and the holographic dark energy by dividing the whole evolution into three pieces, the inflation, the consecutive period of the radiation dominated era and the first half of the matter dominated era, and the consecutive period of the last half of the matter dominated era and the dark energy dominated era.
Identifying the area of the event horizon with the logarithm of the content of missing information, we discuss that the the second law of thermodynamics restricts the value of the constant $d$ to be larger than the square root of the portion of the holographic dark energy $\sqrt{\Omega_{\rm h}}$.
The scale factors at the three equipartition times $t_{\rm eq}$, $t_{\rm md}$, and $t_{\rm h}$ of the matter-radiation, the holographic dark energy-radiation, the matter-holographic dark energy, respectively, are given by
\begin{eqnarray*}
\y(t_{\rm eq})= \frac1{\Of}, \quad \y(t_{\rm md}) \simeq \frac{e^{N_{\rm h}(t_{\rm f})}}{d}, \quad
    \y(t_{\rm h}) = \Of\,e^{2N_{\rm h}(t_{\rm f}) }\frac{c(d)}{d^2}.
\end{eqnarray*}
Since $c(d)$ is an $O(1)$ number, the ratio of two scale factors of neighboring equipartition times are roughly $\Omega_{\rm mf} e^{N_{\rm h}(t_{\rm f})}/d$.
In addition, these three scales provide a very interesting dimensionless constant,
\begin{eqnarray} \label{3a:d}
\frac{\y(t_{\rm eq}) \y(t_{\rm h})}{\y^2(t_{\rm md})} =c(d)=
    \left(\frac{(1+\sqrt{2})^{d}}{2^{d^2/4} (1+\sqrt{2}/d)^{2}}\right)^{\frac{4}{4-d^2}} .
\end{eqnarray}
Note that $c(d)$ is a non-decreasing function of $d$ of order of unity and is independent of all other physical parameters.
Since the three scales will be measurable from experiments, the value of $d$ can be determined from Eq.~(\ref{3a:d}).
Interestingly, the scale factor at the time of full matter dominance is roughly given by the geometric average of the scale factors at the two other equipartition times since $c(d)\sim 1$.
Independently, we have presented another way which determine the value of $d$ by  measuring the deceleration parameter $q$ in Eq.~(\ref{ddR}).

In the presence of a holographic dark energy, the universe must go into the dark energy dominant era eventually.
It is natural to ask why the transition to the dark energy dominant era should happen.
From the point of view of the energy, it is because that the rate of changes of the energy densities are different for each components of the energies, the matter, the radiation, and the holographic dark energy as in Eq.~(\ref{rho:t}).
If the holographic dark energy decreases slower than other densities, it will determine the final fate of the universe.
We have restricted $d\geq 1$ which gives the condition for the velocity of the distance to the horizon, $\dot R_{\rm h} \geq 0$.
This condition, in fact, determines when DDE begins.  
A convincing evidence to this is given by comparing Eq.~(\ref{tmax}) and Eq.~(\ref{Rth}).
The maximum value of the scale factor $\y(t_{\rm max})$ determined from the condition $\dot R_{\rm h} \geq 0$ in phase II, is almost the same as the scale factor $\y(t_{\rm h})$ at which the transition to DDE really happens.

In Ref.~\cite{Davies}, the generalized second law of thermodynamics was studied with the quasi-de Sitter space filled with a viscous fluid in Einstein gravity with a cosmological constant.
Interestingly, they showed that there is a process in which the decrease of the horizon area is supplemented by the increase of the matter entropy to satisfy the generalized second law.
It is an interesting question to ask whether this process is possible or not in the presence of a holographic dark energy.
In the absence of such process, the cosmological arrow of time  becomes the same as the thermodynamical one because of the entropy interpretation of the horizon area.

\begin{acknowledgments}
This work was supported by SRC Program of the KOSEF through the CQUEST grant R11-2005-021 (H.-C. K.) and the Korea Research Foundation Grant funded by Korea Government(MOEHRD, Basic Research Promotion Fund)" (KRF-2006-312-C00095;J.J.L.).
\end{acknowledgments} \vspace{3cm}


\end{document}